\documentclass[11pt]{article}
\usepackage[english]{babel}
\usepackage{latexsym,amsmath,amssymb,amsbsy,amstext,amscd,amsfonts,amsthm}
\usepackage{graphics,graphicx}
\usepackage{color}
\usepackage{tabularx}
\usepackage{multirow}
\usepackage{booktabs}
\usepackage{float}
\theoremstyle{definition}

\usepackage{array}
\usepackage{natbib}
\newcommand{\beq}{\begin{equation}}
\newcommand{\eeq}{\end{equation}}

\newcommand{\beqar}{\begin{eqnarray}}
\newcommand{\eeqar}{\end{eqnarray}}
\newcommand{\bit}{\begin{itemize}}
\newcommand{\eit}{\end{itemize}}
\newcommand{\benum}{\begin{enumerate}}
\newcommand{\eenum}{\end{enumerate}}
\newcommand{\barr}{\begin{array}}
\newcommand{\earr}{\end{array}}

\def\XXint#1#2#3{{\setbox0=\hbox{$#1{#2#3}{\int}$}
   \vcenter{\hbox{$#2#3$}}\kern-.5\wd0}}

\def\b0{\mbox{\boldmath $0$}}

\def\be{\mbox{\boldmath $e$}}

\def\bn{\mbox{\boldmath $n$}}

\newcommand{\bsigma}{\mbox{\boldmath $\sigma$}}

\newcommand{\bvarepsilon}{\mbox{\boldmath $\varepsilon$}}

\def\f0{\ensuremath{\mathbb{O}}}

\date{\today}

\title{Redirection of a crack driven by viscous fluid}

\author{Monika Perkowska$^{(1)}$, Andrea Piccolroaz$^{(2)}$, Michal Wrobel$^{(3)}$ and Gennady Mishuris$^{(4)}$
\\
{\it $^{(1)}$\! EnginSoft S.p.A., Via della Stazione 27, frazione Mattarello, }
\\ {\it 38123 Trento, Italy}
\\
{\it $^{(2)}$Dipartimento di Ingegneria Civile, Ambientale e Meccanica, Universita di Trento,}
\\ {\it via Mesiano, 77 I-38123 Trento, Italy}
\\
{\it $^{(3)}$Faculty of Energy and Fuels, AGH-University of Science and Technology,}
\\{\it 30 Mickiewicza Avenue, 30-059 Krakow, Poland}
\\
{\it $^{(4)}$Department of Mathematics, Aberystwyth University, }
\\ {\it Ceredigion SY23 3BZ, Wales, UK}
}

\begin{document}

\maketitle

\begin{abstract}
As shown by \cite{wrobel_2017}, the hydraulically induced tangential traction on fracture walls changes local  displacement and stress fields. This resulted in the formulation of a new hydraulic fracture (HF) propagation condition based on the critical value of the energy release rate that accounts for the hydraulically-induced shear stress. Therefore it is clear that the crack direction criteria, which depend on the tip distributions of the stress and strain fields, need to be changed. We analyse the two commonly used criteria, one based on the maximum circumferential stress (MCS) and another - on the minimum strain energy density (MSED). We show that the impact of the hydraulically induced shear stress on the direction of the crack propagation is negligible in the case of large material resistance to fracture, while for small toughness the effect is significant. Moreover, values of the redirection angles, corresponding to the so-called viscosity dominated regime ($K_{IC}\to0$), depend dramatically on the ratios of the stress intensity factors.
\end{abstract}


\section{Introduction}

 In the standard approach of Linear Elastic Fracture Mechanics (LEFM), the onset of crack propagation is found by using the energy release rate (ERR) criterion which, in the case of an isotropic elastic material, assumes the form \citep{rice_1968}:
\begin{equation}
\label{eq:ERR1}
{\cal E} =\frac{1+\nu}{E} \left[ (1-\nu)\left( K_I^2 + K_{II}^2 \right) + K_{III}^2\right]={\cal E}_C\equiv \frac{1-\nu^2}{E}K^2_{IC},
\end{equation}
where $\nu$ is the Poisson's ratio and $E$ is the Young's modulus, while ${\cal E}_C$ and $K_{IC}$ are the experimentally found critical values of ERR and material toughness, respectively. Here $K_I$, $K_{II}$ and $K_{III}$ are the stress intensity factors pertaining to three basic modes of fracture load. For pure Mode I loading, equation \eqref{eq:ERR1} transforms into the well known Irwin criterion for crack propagation \citep{Irwin_1957}:
\begin{equation}
\label{Irwin}
K_I = K_{IC}.
\end{equation}
However, for the mixed mode loading, determination of the direction of the crack growth is of crucial importance.

The path of possible crack kinks has been extensively studied for many years (see  \citet{Cotterell_Rice_1980}, \citet{Leblond1989}). Most of the developed theories are based on the information from the Irwin-Williams
expansion of the crack tip field \citep{williams_1957}. Some more advanced criteria utilise additional material parameters related to the underlying physics or other arguments (size of the process zone, size of the possible kink, and so on).

The collection of criteria for kink initiation developed so far to determine the redirection angle in fracture mechanics is extensive. Beginning with the most popular examples: maximum circumferential stress (MCS) \citep{erdogan_1963} and minimum strain energy density (MSED) \citep{Lieb_sih_1968,sih_1974}, we can list the maximum strain energy release rate (MSERR) criterion \citep{pal_knauss_1972,hussain_1974}, the local symmetry criterion \citep{goldstein_1974}, the maximum dilatational strain energy density (MDSED) criterion \citep{theocaris_T_1982,yehia_Y_1991}, the maximum determinant of the stress tensor criterion \citep{papadopoulos_1988}, the J-criterion \citep{hellen_J_1975}, the vector crack tip displacement criterion \citep{li_1989}, the maximum normal strain criterion \citep{chang_1981}, the maximum potential energy release rate criterion \citep{chang_2006}, the so-called T-stresses criteria \citep{williams_1984}, and many others. Clearly, the applicability of any specific approach should be justified on a case by case basis, using the strength properties of the materials involved in the study, the loading conditions and available experimental data to validate the selection of criterion. It follows that there is no universal criterion valid for all possible applications. However, in many situations the discrepancies in prediction given by the different criteria are not large and are usually observable only in the deviation from the pure Mode I load (especially for the infinitesimal kinks most of the criteria coincide - see \citet{Cotterell_Rice_1980}).

When considering hydraulic fracture (HF), the prediction of the possible crack propagation path becomes even more challenging, as the interaction between the pressurised fluid and the solid and complected fracture network substantially increases the complexity of the problem \citep{pal_zim_2017_a,sal_2017}.
Moreover, the sets of credible data that could be used to verify theoretical models are limited or inaccessible. There have also been arguments that cast doubt on the applicability of some of the fracture criteria when applied to brittle fracture \citep{Chudnovsky_Gorelik_1996} and hydraulic fracture \citep{cherny_2017}).

\cite{wrobel_2017} introduced a modified formulation of the HF problem, accounting for a hydraulically induced tangential (asymmetrical) traction at the crack faces. It was shown that, due to the order of the tip singularity of the hydraulic shear stress, this component of the load cannot be omitted when computing ERR. A new parameter, the hydraulic shear stress intensity factor ($K_f$), was introduced and proved to play an important role in the HF process.
The amended crack propagation criterion, under remote Mode I loading conditions, was formulated as:
\begin{equation}
\label{eq:ERR2}
{\cal E}=\frac{1-\nu^2}{E}\left[ K_I^2+ 4(1-\nu) K_IK_f\right]={\cal E}_C.
\end{equation}
This formula includes both, the standard stress intensity factor for Mode I, $K_I$, and the newly introduced hydraulic shear stress intensity factor, $K_f$.

Here we analyse how the shear stress induced by moving fluid at the crack faces influences the crack propagation direction in the most general case, when all fracture modes (Mode I, II, III) are taken into account. We focus on two commonly used criteria, Maximum Circumferential Stress (MCS) and Minimum Strain Energy Density (MSED). Presently, we could not find any experimental data to verify the results and therefore determine which of the two criteria is more relevant to hydraulic fracture problems.

The structure of this paper is as follows. In Section \ref{sec:ERR} a methodology for the computation of the ERR in presence of the hydraulically induced shear stress for mixed mode loading is presented. An asymptotic representation of the stress and strain fields in the vicinity of the fracture tip is given.  In Section \ref{sec:angle}, in a new setting, two criteria are chosen for use in determining the crack propagation angle in the presence of hydraulic tangential traction. Corresponding results are analysed with respect to various crack propagation regimes and values of the Poisson's ratio, and are compared with one another. Finally, we summarise our conclusions in Section \ref{sec:conclusions}.

\section{Computation of the Energy Release Rate accounting for the shear stress induced by fluid in a mixed mode setting}
\label{sec:ERR}

\begin{figure}[htb!]
    \center
    \includegraphics[scale=0.60]{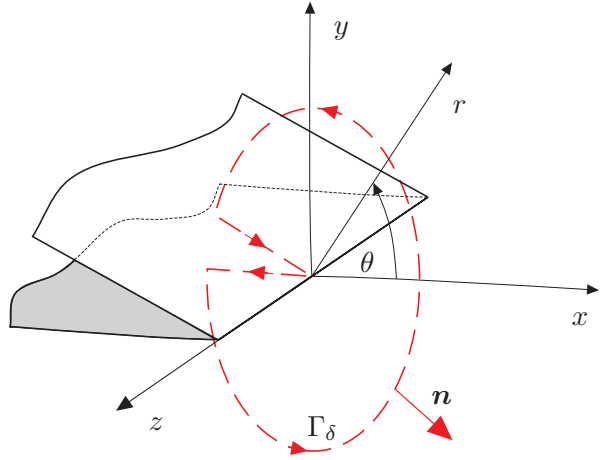}
    \put(-170,10){$z$}
    \put(-10,50){$x$}
    \put(-100,160){$y$}
    \put(-90,70){$\theta$}
    \put(-55,130){$r$}
    \put(-63,20){$\bn$}
    \put(-110,8){$\Gamma_\delta$}
\caption{A planar crack and its local coordinate system.}
\label{crack}
\end{figure}

In the framework of the LEFM, the ERR is computed using the standard $J$-integral argument \citep{rice_1968}:
\begin{equation}
\label{eq:ERR3}
{\cal E}(z) = \lim_{\delta \to 0} J_x^\delta (z) = \lim_{\delta \to 0} \int_{\Gamma_\delta} \Big\{ \frac{1}{2} ({\boldsymbol \sigma} \cdot {\boldsymbol \varepsilon}) n_x - {\boldsymbol t}_n \cdot \frac{\partial {\boldsymbol u}}{\partial x} \Big\} ds,
\end{equation}
where $\Gamma_\delta$ is a circular contour of radius $\delta$ around the fracture tip, contained in a plane orthogonal to the crack front, $\bf n$ is the outward normal to the contour $\Gamma_\delta$, and ${\bf t}_n = {\boldsymbol \sigma}\bf n$ is the traction vector along $\Gamma_\delta$ (see Fig.~\ref{crack}).

%%%%%%%%%%%%%%%%%%%%%%%%%%%%%%%%%%%%%%%%%%%%%%%%%%%%%%%%%%%%%%%%%%%%%%%%
%%%%%%%%%%%%%%%%%%%%%%%%%%%%%%%%%%%%%%%%%%%%%%%%%%%%%%%%%%%%%%%%%%%%%%%%

\begin{figure}[htb!]
\center
    \includegraphics [scale=0.70]{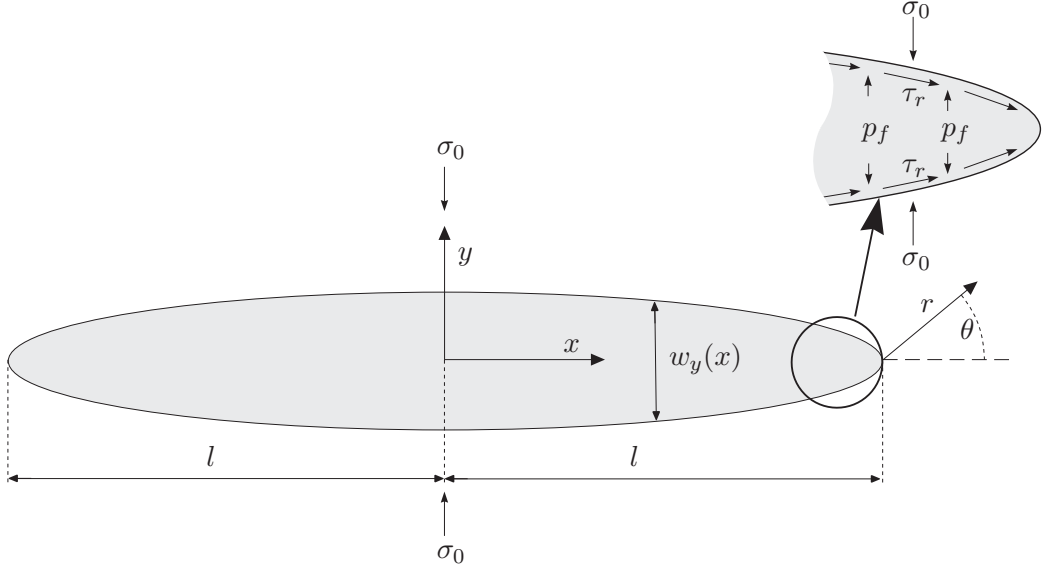}
    \put(-180,70){$x$}
    \put(-220,105){$y$}
    \put(-140,65){$w_y(x)$}
    \put(-155,27){$l$}
    \put(-315,27){$l$}
    \put(-30,75){$\theta$}
    \put(-45,85){$r$}
    \put(-67,152){$p_f$}
    \put(-37,152){$p_f$}
    \put(-52,165){$\tau_r$}
    \put(-52,138){$\tau_r$}
    \put(-228,145){$\sigma_0$}
    \put(-228,-8){$\sigma_0$}
    \put(-52,198){$\sigma_0$}
    \put(-52,103){$\sigma_0$}
\caption{Sketch of a plane-strain fluid driven fracture. }
\label{figure_c}
\end{figure}

The classical fracture criterion \eqref{eq:ERR1} is derived directly from formula \eqref{eq:ERR3}, for an arbitrary mixed mode deformation and smooth crack front. It has been widely adopted in the analysis of hydraulic fracture \citep{garagash_det_1999,bunger_2005,gar_2006,ad_et_al_2007,wrobel_2015,Perkowska_2016} on the ad hoc assumption that the hydraulically induced tangential traction is small compared to the net fluid pressure and can thus be neglected. However, \cite{wrobel_2017} showed that the singularity of the hydraulic shear stress is stronger than that of the fluid pressure, and therefore the former cannot be omitted when deriving the integrals in \eqref{eq:ERR3}. Indeed, in accordance with lubrication theory, the shear stress acting on the crack faces can be computed as (see e.g. \cite{batchelor_1976}):
\begin{equation}
\label{eq:lub}
{\boldsymbol \tau}(r,\theta,z)_{|_{\theta=\pm\pi}} =
\mp\frac{w_y(r,z)}{2} \nabla_{(r,z)} p(r,z) =
\mp\frac{1}{2} w_y\left[ \frac{\partial p}{\partial r} \,\be_r + \frac{\partial p}{\partial z} \, \be_z \right],
\end{equation}
where $w_y(r,z)$ is the width of the crack opening in the direction orthogonal to the crack faces, $p(r,z)=p_f(r,z)-\sigma_0$ is the so-called net fluid pressure in the channel ($p_f$ - fluid pressure, see Fig.~\ref{figure_c}), and $\be_r, \be_z$ denote the respective unit vectors.

When considering the tip asymptotics of HF in the so-called toughness dominated regime, which was proved in \cite{wrobel_2017} to be the only permissible type of solution behaviour in the vicinity of the fracture front, we obtain the following estimates for Mode I:
\begin{equation}
\label{eq:wp}
w_y(r,z)=w_0(z)\sqrt{r}+O(r),\quad p(r,z)=p_0(z)\log(r)+O(1), \quad r \to 0,
\end{equation}
and
\begin{equation}
\label{eq:lubas}
\tau_r(r,\theta,z)_{|_{\theta=\pm\pi}}=\mp\frac{\tau_0(z)}{\sqrt{r}}+O(1),  \quad r \to 0,
\end{equation}
where the multipliers of the leading asymptotic terms are interrelated: %
\begin{equation}
\label{eq:tau0}
\tau_0(z)=\frac{1}{2}p_0(z) w_0(z).
\end{equation}
%

%%%%%%%%%%%%%%%%%%%%%%%%%%%%%%%%%%%%%%%%%%%%%%%%%%%%%%%%%%%%%%%%%%%%%%%%
%%%%%%%%%%%%%%%%%%%%%%%%%%%%%%%%%%%%%%%%%%%%%%%%%%%%%%%%%%%%%%%%%%%%%%%%

In the general case, the complete asymptotic expansions of the displacement and stress fields in the near-tip region ($r\to0$) are:
\begin{equation}
\label{eq:asymu}
{\bf u}(r,\theta,z)=\sqrt{\frac{r}{2\pi}}\Big[ K_I{\bf \Phi}_{I}(\theta)+K_{II}{\bf\Phi}_{II}(\theta)+K_{III}{\bf\Phi}_{III}(\theta)+K_f{\bf \Phi}_{\tau}(\theta)
\Big]+O\left(r \log r\right),
\end{equation}
\begin{equation}
\label{eq:asyms}
{\boldsymbol \sigma}(r,\theta,z)=\frac{1}{\sqrt{2\pi r}}\left[ K_I{\bf \Psi}_{I}(\theta)+K_{II}{\bf\Psi}_{II}(\theta)+K_{III}{\bf\Psi}_{III}(\theta)+K_f{\bf\Psi}_{\tau}(\theta)\right]+O\left(\log r\right),
\end{equation}
where $\{r,\theta,z\}$ is a local polar coordinate system (see Fig.~\ref{crack}), $K_I$, $K_{II}$ and $K_{III}$ are the classical stress intensity factors (SIFs) and $K_f$ is the hydraulic shear stress intensity factor (HSSIF) related to the hydraulic shear stress, $\tau$ (see \eqref{eq:lubas}). The functions ${\bf \Phi}_j(\theta)$ and ${\bf \Psi}_j(\theta)$ define the polar angle dependence and are given in Appendix \ref{app1}.
%Note that ${\bf u}=[u_r,u_{\theta}]$ and  $w_y(r,z)=u_{\theta}(r,-\pi,z)-u_{\theta}(r,\pi,z)$.
Clearly, all the stress intensity factors in the asymptotic relationships \eqref{eq:asymu} -- \eqref{eq:asyms} depend on $z$, while the vector-functions ${\bf \Phi}_j$ and ${\bf \Psi}_j$ are $z$-independent. Since we will only analyse the local (2D) problem in this paper, we will omit the $z$ variable henceforth.

The above representations were constructed as a superposition of four displacement and stress fields, three of which are related to the classical fracture mechanics loads (Mode I, II and III, where the traction vanishes at the crack surfaces) and the fourth, which is a result of the hydraulic action of fluid via the shear stresses induced on the crack faces.

The asymptotics corresponding respectively to the components of the crack opening, $w_j(r)=u_j(r,+\pi)-u_j(r,-\pi)$, can be expressed by:
\begin{equation}
\label{eq:wy}
w_y(r) =\gamma(K_I+K_f)\sqrt{r}+O(r),\quad r\to0,
\end{equation}
\begin{equation}
\label{eq:wxz}
w_x(r) =\gamma K_{II}\sqrt{r}+
O(r^{3/2}),\quad
w_z(r) =\gamma K_{III}\sqrt{r}+O(r^{3/2}),\quad r\to0.
\end{equation}

Comparing these with \eqref{eq:wp} gives:
\begin{equation}
\label{eq:w0}
w_0=\gamma(K_I+K_f),\quad
K_f=\sqrt{\frac{\pi}{2}} \frac{\tau_0}{1-\nu},\quad \gamma=\frac{8}{\sqrt{2\pi} }\frac{1-\nu^2}{E}.
\end{equation}

We note that $K_I$ and $K_f$ are not independent. Indeed, combining \eqref{eq:tau0} and \eqref{eq:w0} we find:

\begin{equation}
\label{eq:Kf}
K_f=\varpi K_I,\quad \varpi=\frac{p_0}{G-p_0}>0,
\end{equation}
where $G=\frac{E}{2(1+\nu)}$ is the shear modulus of the elastic material, and the dimensionless parameter $\varpi$ varies from $0$ to $\infty$. For more details, see \cite{wrobel_2017}.

A new formula for the ERR, following from \eqref{eq:ERR3} and \eqref{eq:asymu} -- \eqref{eq:asyms}, can now be given as:
\begin{equation}
\label{eq:ERRnew}
{\cal E}=\frac{1+\nu}{E} \left\{ (1-\nu)\left[ K_I^2 + K_{II}^2 + 4(1-\nu) K_IK_f\right] + K_{III}^2 \right\},
\end{equation}
which leads to the fracture criterion:
\begin{equation}
\label{eq:ERRnew2}
K_I^2 + K_{II}^2 + 4(1-\nu) K_I K_f + \frac{1}{1-\nu} K_{III}^2 =K_{IC}^2.
\end{equation}
We note that conditions \eqref{eq:ERR1} and \eqref{eq:ERR2} are particular forms of the general formula \eqref{eq:ERRnew2}. When the stress intensity factors $K_{II}$ and $K_{III}$ are defined by the external conditions, the expression:
\begin{equation}
\label{K_I_eff}
K_{IC}^{\,e\!f\!f}=\sqrt{K_{IC}^2- K_{II}^2-\frac{1}{1-\nu} K_{III}^2}
\end{equation}
can be considered to be an ``effective toughness'' for the hydraulic fracture problem under the mixed load.

%In case then the stress intensity factors $K_{II}$ and $K_{III}$ are defined from the elasticity problem only (shear stress coming from the external load
%for crack of special geometry), the expression $K_{IC}^{\,e\!f\!f}=\sqrt{K_{IC}^2- K_{II}^2-\frac{1}{1-\nu} K_{III}^2}$ represents the "effective toughness" when hydraulic fracture propagation regime is in question.

\subsection{Normalisation}
\label{sec:norm}

In order to facilitate the parametric study, we now introduce the following natural scaling:
\begin{equation}
\label{eq:Knorm}
\hat K_I =\frac{K_I}{K_{IC}},\quad \hat K_{II} =\frac{K_{II}}{K_{IC}},\quad \hat K_{III} =\frac{K_{III}}{K_{IC}},\quad \hat K_f=\frac{K_f}{K_{IC}}.
\end{equation}
The fracture criterion \eqref{eq:ERRnew2} now becomes:
\begin{equation}
\label{eq:ERRnorm}
\hat K_I^2 + \hat K_{II}^2 + 4(1-\nu) \hat K_I \hat K_f + \frac{1}{1-\nu} \hat K_{III}^2 =1.
\end{equation}
We note that, under such a normalization, the material resistance to brittle fracture described by $K_{IC}$,  is introduced implicitly. On the other hand, identification of the crack propagation regime (viscosity dominated, small and large toughness modes) hinges on this property. Thus  we introduce a dimensionless parameter $\tilde p_0=2\pi p_0(1-\nu^2)/E$, related to material toughness (compare with equation $(71)$ in the work of \cite{wrobel_2017}), that combines the stress intensity factors $\hat K_I$ and $\hat K_f$ in the same manner as in equation \eqref{eq:Kf}:
\begin{equation}
\label{eq:Kfnorm}
 \varpi=\frac{\tilde p_0}{\pi(1-\nu)-\tilde p_0}, \quad \hat K_f=\varpi \hat K_I,\quad 0<\tilde p_0<\pi(1-\nu).
\end{equation}

The values of the parameter $\tilde p_0$ and the stress intensity factors are not independent. As was shown by \cite{wrobel_2017} for the Mode I deformation ($K_{II}=K_{III}=0$), parameter $\tilde p_0$ determines the crack propagation regime ($\tilde p_0\to0$ corresponds to the toughness dominated one, while $\tilde p_0\to \pi(1-\nu)$ defines the viscosity dominated mode). Taking \eqref{K_I_eff} into account, we conclude that:
\begin{equation}
\label{nob_mode1_p0}
K_{IC}\cdot \hat K^{\,e\!f\!f}_{IC}\to\infty \quad  \Leftrightarrow  \quad
\hat K_I \to 1\quad \mbox{and}\quad  \tilde p_0 \to 0,
\end{equation}
and
\begin{equation}
\label{nob_mode1_p01}
K_{IC}\cdot \hat K^{\,e\!f\!f}_{IC}\to0 \quad  \Leftrightarrow \quad
\hat K_I \to 0  \quad \text{and} \quad \tilde p_0 \to \pi(1-\nu),
\end{equation}
where the normalised effective toughness is defined as follows:
\begin{equation}
%\label{nob_mode1_p01}
\hat K^{\,e\!f\!f}_{IC}=\sqrt{1-\hat K_{II}^2-\frac{1}{1-\nu}\hat K_{III}^2}.
\end{equation}

The combination of \eqref{eq:ERRnorm} and \eqref{eq:Kfnorm} yields, after rearrangement, a formula for $\hat K_I$, provided that $\hat K_{II}$,  $\hat K_{III}$ and $\tilde p_0$ are known:
\begin{equation}
\label{eq:KInorm}
\hat K_I=\sqrt{\frac{\pi(1-\nu)-\tilde p_0}{\pi(1-\nu)+\tilde p_0(3-4\nu)}}\,\hat K^{\,e\!f\!f}_{IC}.
\end{equation}

After substitution of \eqref{eq:Kfnorm} into \eqref{eq:KInorm}, and some algebra, we obtain a relation for $\hat K_f$:
\begin{equation}
\label{eq:Kform2}
\hat K_f=\frac{\tilde p_0}{\sqrt{\big[\pi(1-\nu)-\tilde p_0\big]\big[\pi(1-\nu)+\tilde p_0(3-4\nu)\big]} }\,\hat  K^{\,e\!f\!f}_{IC}.
\end{equation}
It can be easily seen that for any fixed values of $\hat K_{II}$ and $\hat K_{III}$ we have, when comparing with \eqref{eq:KInorm} and \eqref{eq:Kform2}:
\begin{equation}
\label{eq:limK}
\lim_{\tilde p_0\to\pi(1-\nu)}\hat K_f\hat K_I=\frac{1}{4(1-\nu)}\Big(\hat K^{\,e\!f\!f}_{IC}\big)^2.
\end{equation}
Finally, we note that for $\hat K_{III}=0$ and $\tilde p_0=0$ (classical mixed Mode I and II), both normalised stress intensity factors, $\hat K_I=\sqrt{1-\hat K_{II}^2}$ and $\hat K_f=0$, are independent of $\nu$.

Equations \eqref{eq:KInorm} and \eqref{eq:Kform2} provide a relationship between the normalised symmetric SIFs, $\hat K_I$ and $\hat K_f$, and the normalised anti-symmetric SIFs, $\hat K_{II}$ and $\hat K_{III}$, while also taking into account the influence of the hydraulically induced shear stresses through the pressure parameter $\tilde{p}_0$. This allows for a parametric analysis of the fracture propagation angle, where the independent parameters are $\hat K_{II}$, $\hat K_{III}$ and $\tilde{p}_0$. This analysis is given in the next section.

\section{Determination of the fracture propagation angle}
\label{sec:angle}

If the crack is only under a Mode I load ($\hat K_{II}=\hat K_{III}=0$), it propagates in a self-similar fashion (so that the fracture propagation angle $\theta_f$ is equal to zero). However, fractures are often subjected to mixed-mode loadings \citep{qian_1996}. Therefore, an accurate prediction of the fracture orientation is crucial to defining the path of a crack kink.

In the analysis below, the propagation angle, $\theta_f$, will be determined in the most general case, when all three modes are present. However, when examining the influence of the hydraulically induced shear stress, the analysis will be restricted to the case of mixed Mode I and II ($K_{III}=0$). In fact, most fractures in geological formations occur under such loading \citep{li_2013}. On the other hand, the applicability of the classical crack redirection criteria raises doubts when accounting for the impact of severe Mode III loading \citep{lazarus_2008}. Recently, an attempt has been made to tackle such cases \citep{cherny_2016,cherny_2017}.

\subsection{Maximum Circumferential Stress (MCS) criterion}
\label{sec:mcs}

The MCS criterion was introduced by \cite{erdogan_1963}. It states that the crack will propagate in the direction where the hoop stress $\sigma_{\theta\theta}$ reaches its maximum over the interval $-\pi<\theta<\pi$:
\begin{equation}
\label{eq:mcs}
\theta_f=\theta \Big|_{\sigma_{\theta\theta}=\sigma_{\theta\theta}^\text{max}}.
\end{equation}
Taking \eqref{eq:asyms} into account, we have:
\begin{equation}
\label{eq:sigtt}
\sigma_{\theta\theta}(r,\theta)=\frac{K_{IC}}{\sqrt{2 \pi r}}\left[ \hat K_I\cos^3\frac{\theta}{2}-3\hat  K_{II}\sin\frac{\theta}{2}\cos^2\frac{\theta}{2}+2(1-\nu) \hat K_f\cos\frac{3\theta}{2} \right].
\end{equation}
According to this formula, the direction of the crack propagation does not depend on the Mode III component or the material toughness $K_{IC}$. Instead, it hinges on the relationship between the normalised stress intensity factors $\hat K_I$, $\hat K_{II}$, $\hat K_f$ and the Poisson's ratio, $\nu$, if $\hat K_f>0$. It is important to note that there exists only one value of $\theta_f$ in the interval $-\pi<\theta<\pi$ that satisfies equation \eqref{eq:sigtt}.

\begin{figure}[htb!]
\begin{center}
\includegraphics[scale=0.5]{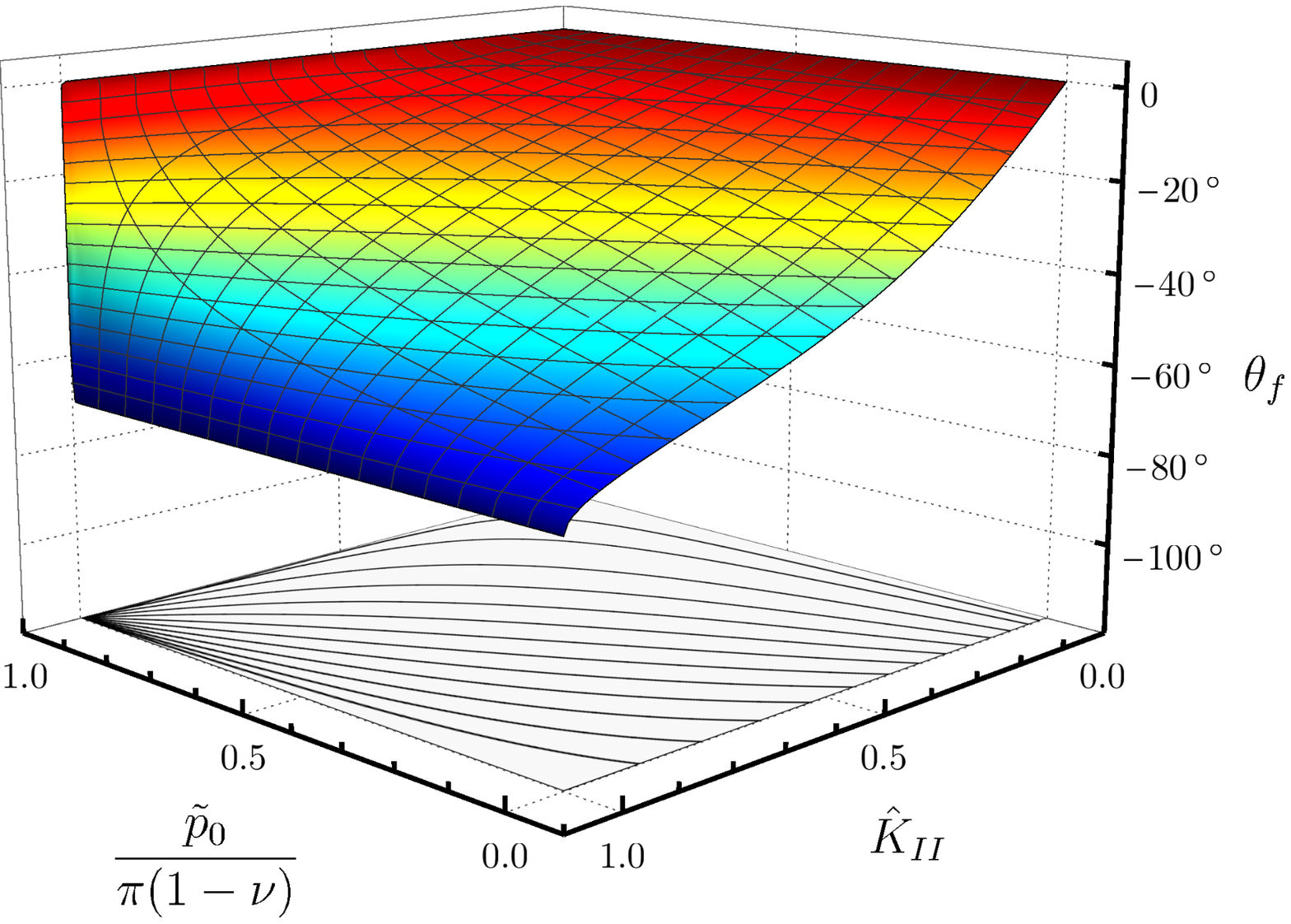}
\hspace{0mm}
\put(-150,60){$D$}
\put(-150,190){$B$}
\put(-245,180){$C$}
\put(-50,180){$A$}
\includegraphics[scale=0.45]{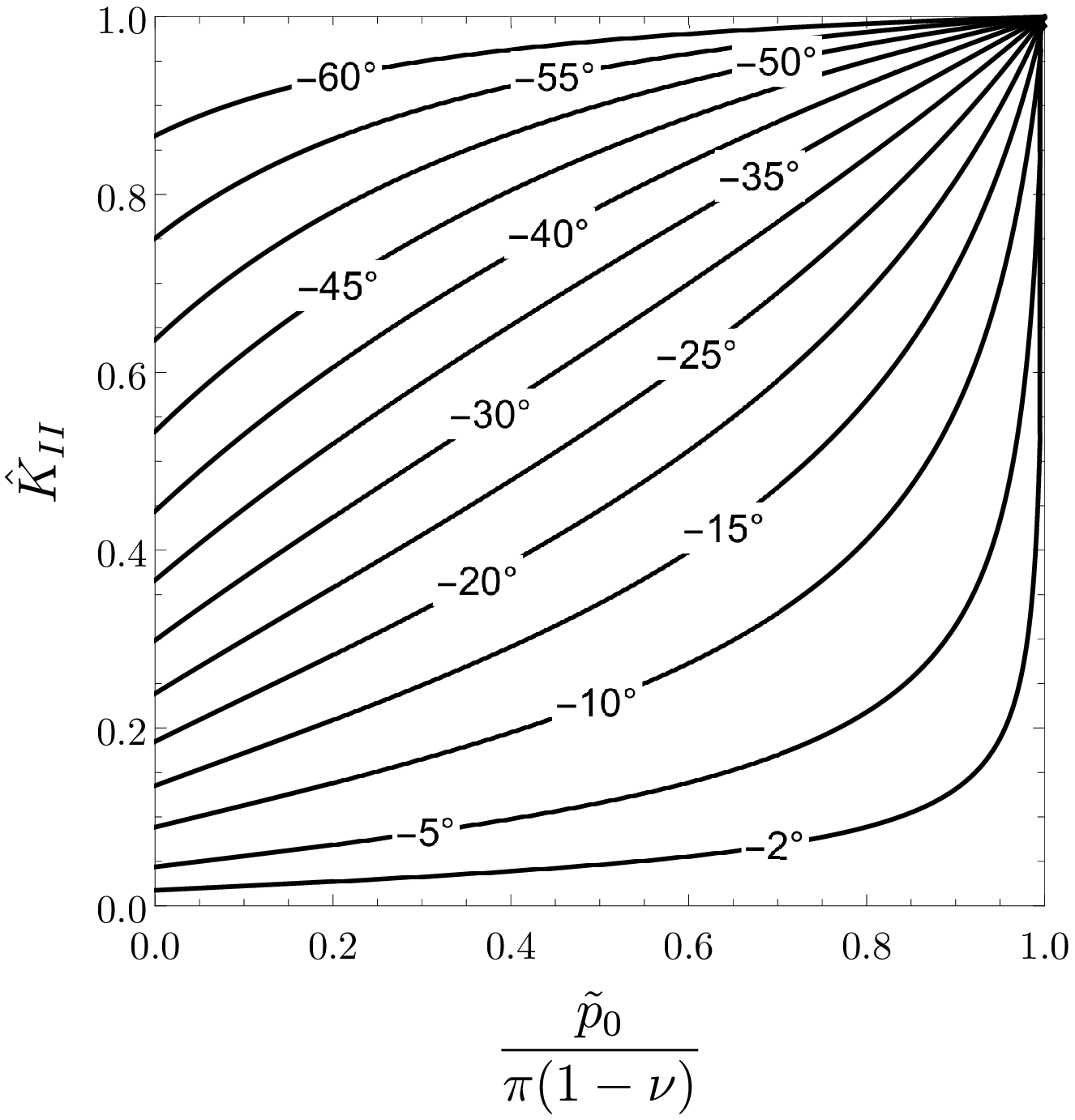}
\put(-165,20){$A$}
\put(5,20){$B$}
\put(-165,180){$D$}
\put(5,180){$C$}
\caption{MCS: Predicted propagation angle $\theta_f$ for $\hat K_{II} \in [0,1]$ and $\tilde p_0 \in [0,\pi(1-\nu)]$ for $\nu=0.3$. }
\label{MCS_n03}
\end{center}
\end{figure}

The fracture propagation angle $\theta_f$ computed according to the MCS criterion \eqref{eq:mcs} is presented in Fig.~\ref{MCS_n03} for  $\nu=0.3$, and for all admissible values of $\hat K_{II} \in [0,1]$ and $\tilde p_0 \in [0,\pi(1-\nu)]$.

%%%%%%%%%%%%%%%%%%%%%%%%%%%%%%%%%%%%%%%%%%%%%%%%%%%%%%%%%%%%%%%%%%%%%%%%
%%%%%%%%%%%%%%%%%%%%%%%%%%%%%%%%%%%%%%%%%%%%%%%%%%%%%%%%%%%%%%%%%%%%%%%%

The limiting regimes, defined in the parametric space in which the fracture evolves, are denoted by vertices $A, B, C, D$ in Fig.~\ref{MCS_n03}b.

The value of $\theta_f$ in the case of classical linear elastic fracture mechanics ($\tilde p_0=0$ or edge $AD$) was found analytically (see \cite{erdogan_1963}):
\begin{equation}
\label{mcs_analytical}
\theta_f=2 \arctan \left(\frac{\hat K_I}{4\hat K_{II}}-\sqrt{\frac{1}{2}+\frac{\hat K_I^2}{16\hat K_{II}^2}} \right).
\end{equation}

On the edge $CD$ ($\hat K_{II}=1$), we have from \eqref{nob_mode1_p01} and \eqref{eq:KInorm} that $K^{\,e\!f\!f}_{IC}=\hat K_I=0$, while the angle $\theta_f$ can also be found from \eqref{mcs_analytical}. This conclusion is, however, not true at the corner C, which also lies on the edge $BC$ corresponding to the so-called viscosity dominated regime ($\tilde p_0\to \pi(1-\nu)$, the amount of energy dissipated in the viscous fluid flow is much greater than that released in the brittle fracture). Here, we have $\theta_f=0$, as the maximum value of the circumferential stress in this case is defined by the third term in \eqref{eq:sigtt}.

To derive the value of $\theta_f$ on the last edge, $AB$ ($\hat K_{II}=0$), we must simultaneously maximise the first and the last terms in \eqref{eq:sigtt}, which holds for $\theta_f=0$.

We note that in the vicinity of corner $C$, we observe high sensitivity in the angle of crack propagation, $\theta_f$, to changes in the parameters $\tilde p_0$ and $\hat K_{II}$. In fact, the function $\theta_f\big(\tilde p_0,\hat K_{II}\big)$ does not possess a limit at point $(1,1)$. Indeed, when using the asymptotic estimate \eqref{eq:limK} and the relationship \eqref{nob_mode1_p01} we have:
\begin{equation}\label{lim_C}
\hat K_f\sim \frac{1-\hat K_{II}^2}{4(1-\nu)\hat K_I}\quad \text{as} \quad \tilde p_0\to\pi(1-\nu),
\end{equation}
where both the numerator and the denominator tend to zero as $\tilde p_0 \to \pi(1-\nu)$. As a result, the ratio describing the coefficient in the third term in \eqref{eq:sigtt} is indeterminate. Therefore, $\theta_f$  depends on the load history, and for this reason the limit of expression \eqref{lim_C} does not exist at point C. Physically, this phenomenon can be explained by the competition between pure Mode II fracture and the viscosity dominated regime of crack propagation (each of these having different propagation angles).

\begin{figure}[htb!]
\begin{center}
\includegraphics[scale=0.7]{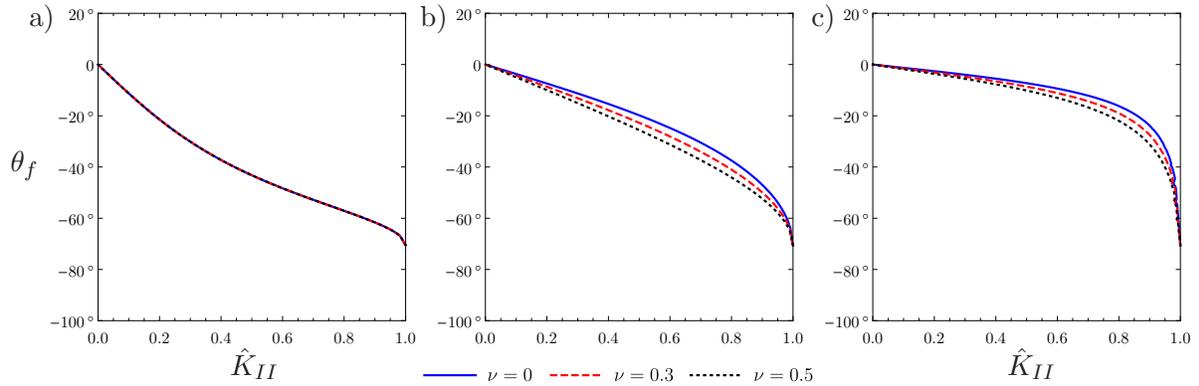}
\put(-447,80){$\theta_f$}
\put(-364,3){$\hat K_{II}$}
\put(-70,3){$\hat K_{II}$}
\put(-440,135){a)}
\put(-292,135){b)}
\put(-144,135){c)}
\caption{MCS: Redirection angle, $\theta_f$, for various values of Poisson's ratio and: a) $\frac{\tilde p_0}{\pi(1-\nu)}=0$, b) $\frac{\tilde p_0}{\pi(1-\nu)}=0.5$, c) $\frac{\tilde p_0}{\pi(1-\nu)}=0.9$.}
\label{MCS_nu}
\end{center}
\end{figure}

Let us now analyse a possible impact of the Poisson's ratio on the direction of crack propagation (Fig. \ref{MCS_nu}). As expected, for the standard mixed-mode case (without accounting for the singular term induced by the fluid that is $\tilde p_0=0$, $\hat K_f=0$), the redirection angle does not depend on the Poisson's ratio $\nu$. This follows immediately from equation \eqref{eq:sigtt} or \eqref{mcs_analytical} and can be seen in Fig. \ref{MCS_nu}a).

%%%%%%%%%%%%%%%%%%%%%%%%%%%%%%%%%%%%%%%%%%%%%%%%%%%%%%%%%%%%%%%%%%%%%%%%
%%%%%%%%%%%%%%%%%%%%%%%%%%%%%%%%%%%%%%%%%%%%%%%%%%%%%%%%%%%%%%%%%%%%%%%%

\begin{figure}[htb!]
\begin{center}
\includegraphics[scale=1.1]{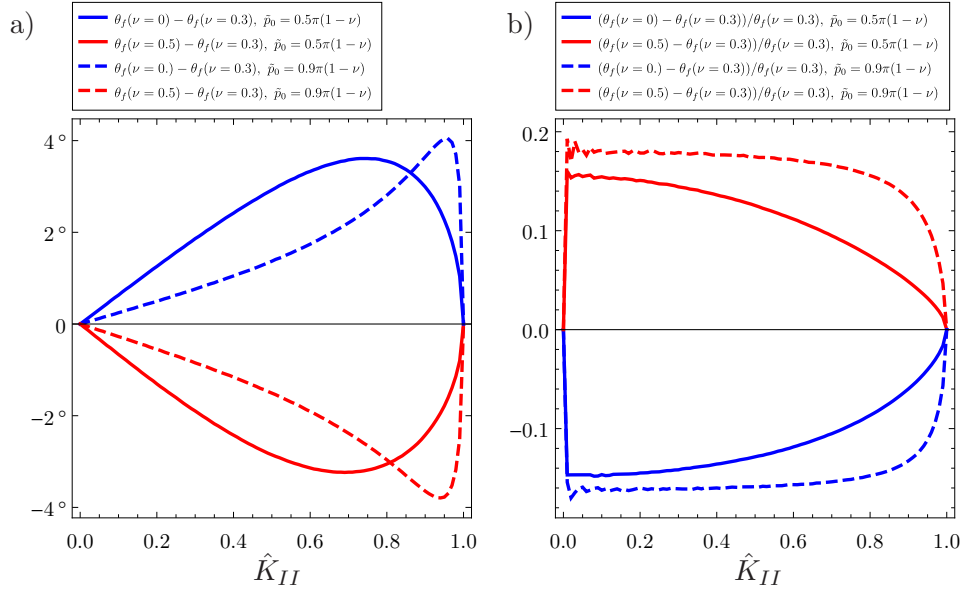}
%\put(-430,66){$\theta_f$}
\put(-268,-12){$\hat K_{II}$}
\put(-86,-12){$\hat K_{II}$}
\put(-360,195){a)}
\put(-172,195){b)}
\caption{MCS: Absolute (a) and relative (b) deviations of the redirection angle, $\theta_f$, from a reference value obtained in the case of $\nu=0.3$ for two limiting values of the Poisson's ratio: $\nu=0$ and $\nu=0.5$.}
\label{MCS_nu_diff}
\end{center}
\end{figure}

In the general case of $\tilde p_0>0$, the impact of the Poisson's ratio is relatively weak and vanishes when approaching the ends of the interval ($\hat K_{II}=0$ and $\hat K_{II}=1$). As $\tilde p_0$ increases, the maximal deviations between respective propagation angles (for various $\nu$) are located closer to the right end of the $\hat K_{II}$ interval (see Fig. \ref{MCS_nu_diff}). However, the differences between the redirection angles for various Poisson's ratios are hardly distinguishable, giving maximal deviations between respective results of up to $4^\circ$ (compare Fig. \ref{MCS_nu_diff}). Thus, according to the MCS criterion, the influence of the Poisson's ratio on crack redirection can be neglected for practical applications, regardless of the fracture propagation regime.

%%%%%%%%%%%%%%%%%%%%%%%%%%%%%%%%%%%%%%%%%%%%%%%%%%%%%%%%%%%%%%%%%%%%%%%%
%%%%%%%%%%%%%%%%%%%%%%%%%%%%%%%%%%%%%%%%%%%%%%%%%%%%%%%%%%%%%%%%%%%%%%%%

\subsection{Minimum Strain Energy Density (MSED) criterion}
\label{sec:msed}

Another popular fracture propagation criterion is based on the minimum strain energy density (MSED) It was  proposed by \cite{Lieb_sih_1968} and \cite{sih_1974}. For the strain energy density:
\begin{equation}\label{strain_energy_density}
W=\frac{1}{2} \bsigma \cdot \bvarepsilon,
\end{equation}
it is assumed that the factor $S=W r$ takes its minimal value in the direction of possible crack propagation:
\begin{equation}
\label{tf_S}
\theta_f=\theta \Big|_{S=S^{\text{min}}}.
\end{equation}
In the setting of the present paper, the factor $S$ is computed as:
\begin{multline}
\label{S_0}
S(\theta)=\frac{1+\nu}{2\pi E}K_{IC}^2\Big\{\frac{\hat K_I^2}{2}\cos^2\frac{\theta}{2}\left[3-4\nu-\cos\theta\right]+\frac{\hat K_{II}^2}{8}\left[9-8\nu-4(1-2\nu)\cos\theta+3\cos 2\theta\right] \\[2mm]
+\hat K_{III}^2 + 4(1-\nu)^2\hat K_f^2 +\hat K_I\hat K_{II}\sin\theta\left[2\nu-1+\cos\theta\right]+(1-\nu)\hat K_f\left(2\hat K_I\sin^2\theta+  \hat K_{II}\sin 2\theta\right)\Big\}.
\end{multline}
We note that according to this criterion, and in contrast to MCS, the value of $\theta_f$ depends on both the stress intensity factor for Mode III, $\hat K_{III}$, and the Poisson's ratio, $\nu$, even in the case of classical LEFM ($\tilde p_0=0$). However, to remain in the same parametric space in our analysis, the Mode III component will be assumed to be zero.

\begin{figure}[htb!]
\begin{center}
\includegraphics[scale=0.7]{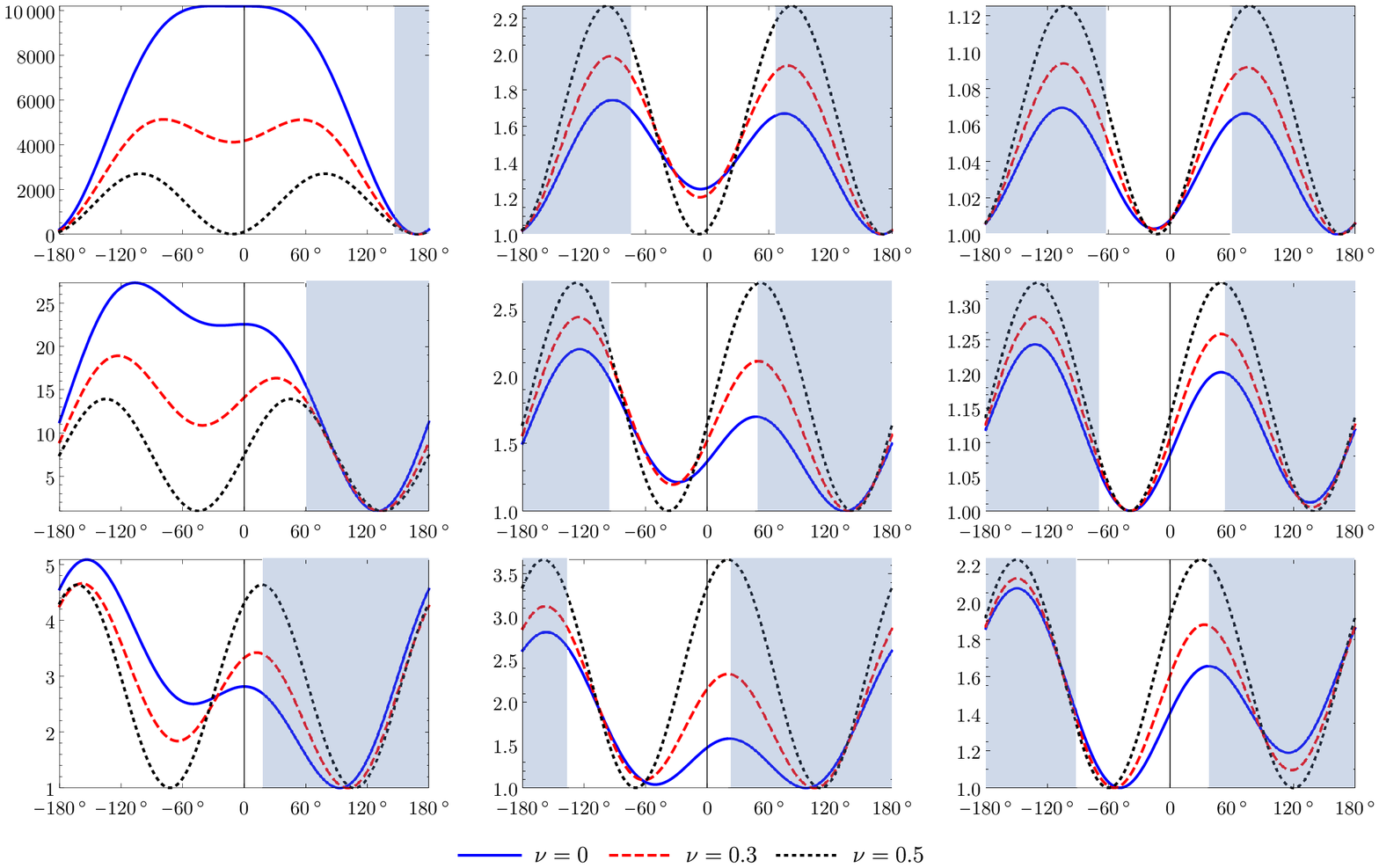}
\put(-455,55){$\frac{S(\theta)}{S^{\text{min}}}$}
\put(-455,145){$\frac{S(\theta)}{S^{\text{min}}}$}
\put(-455,235){$\frac{S(\theta)}{S^{\text{min}}}$}
\put(-361,3){$\theta$}
\put(-68,3){$\theta$}
%\put(-445,266){a)}
%\put(-292,266){b)}
%\put(-147,266){c)}
%\put(-445,178){d)}
%\put(-292,178){e)}
%\put(-147,178){f)}
%\put(-445,90){g)}
%\put(-292,90){h)}
%\put(-147,90){i)}
\put(0,235){$\hat K_{II}=0.1$}
\put(0,145){$\hat K_{II}=0.5$}
\put(0,55){$\hat K_{II}=0.9$}
\put(-389,281){$\frac{\tilde p_0}{\pi(1-\nu)}=0$}
\put(-240,281){$\frac{\tilde p_0}{\pi(1-\nu)}=0.5$}
\put(-98,281){$\frac{\tilde p_0}{\pi(1-\nu)}=0.9$}
%\put(-400,281){$\tilde p_0/(\pi(1-\nu))=0$}
%\put(-255,281){$\tilde p_0/(\pi(1-\nu))=0.5$}
%\put(-111,281){$\tilde p_0/(\pi(1-\nu))=0.9$}

\caption{MSED: Value of $S(\theta)/S^{\text{min}}$ for various values of Poisson's ratio and fixed $\hat K_{II}$ and $\tilde p_0$. The grey regions on the graphs correspond to the areas where $\sigma_{\theta\theta}<0$.}
\label{MSED_func}
\end{center}
\end{figure}

In Fig.~\ref{MSED_func} the graphs of $S/S^{\text{min}}$ are plotted for three values of $\nu=\{0,0.3,0.5\}$. Each graph refers to a fixed value of $\tilde p_0/(\pi(1-\nu))=\{0,0.5,0.9\}$ and a fixed value of $\hat K_{II}=\{0.1,0.5,0.9\}$.

%%%%%%%%%%%%%%%%%%%%%%%%%%%%%%%%%%%%%%%%%%%%%%%%%%%%%%%%%%%%%%%%%%%%%%%%
%%%%%%%%%%%%%%%%%%%%%%%%%%%%%%%%%%%%%%%%%%%%%%%%%%%%%%%%%%%%%%%%%%%%%%%%

In each case, it can be seen that there are two local minima (similar behaviour was observed for the classical LEFM by \cite{sih_macdonald_1974}) that cause ambiguity in identification of the crack redirection angle, that was noticed by \cite{chang_1982}. Moreover, in the work of \cite{swedlow_1976} there were indications that, for many combinations of loading modes, the selection of the global minimum of the strain energy density, which in turn corresponds to a global relative maximum of potential energy, leads to incorrect values of the redirection angle. As a result of this analysis, a modification to Sih's statement was proposed that the sought energy minimum does not need to be global. Instead, a local value that corresponds to a positive tensile circumferential stress can be taken:
\begin{equation}
\label{tf_S_new}
\theta_f=\theta \Big|_{\{S=S^{\text{min}}\}\wedge \{\sigma_{\theta\theta}>0\}}.
\end{equation}
Furthermore, in the work of \cite{baydoun_2012}, it was suggested that ``{\it the smaller absolute angle is considered as the propagation angle}''. We have checked computationally that both assumptions lead to the same result or, in other words, that the minimum of $S$ obtained for the smallest value of $\theta$ is also the one that corresponds to $\sigma_{\theta \theta}>0$. We believe that those assumptions represent a natural choice for the fracture propagation angle according to the MSED criterion.

\begin{figure}[htb!]
\begin{center}
\includegraphics[scale=0.5]{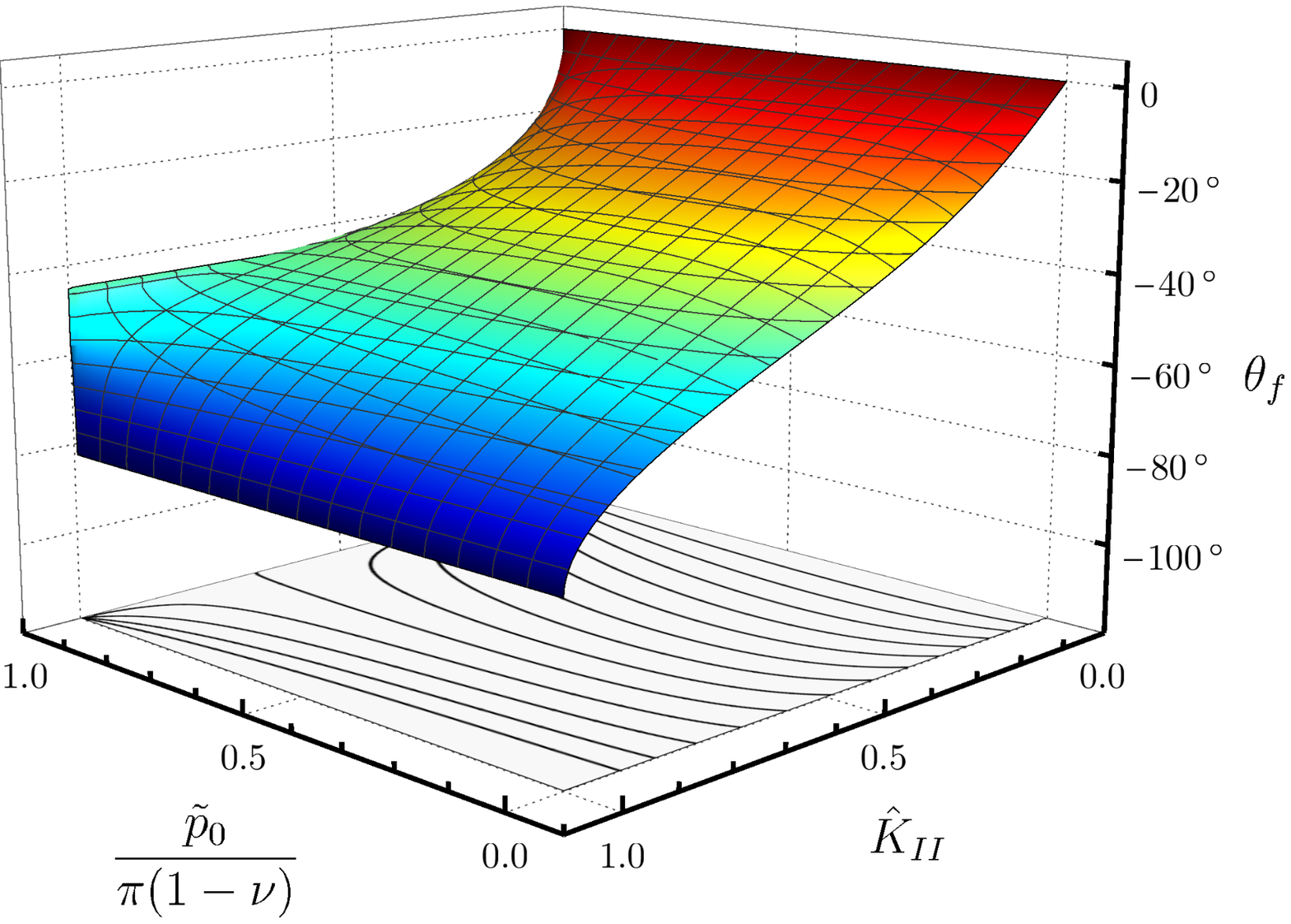}
\hspace{0mm}
\put(-150,50){$D$}
\put(-150,190){$B$}
\put(-240,135){$C$}
\put(-50,180){$A$}
\includegraphics[scale=0.45]{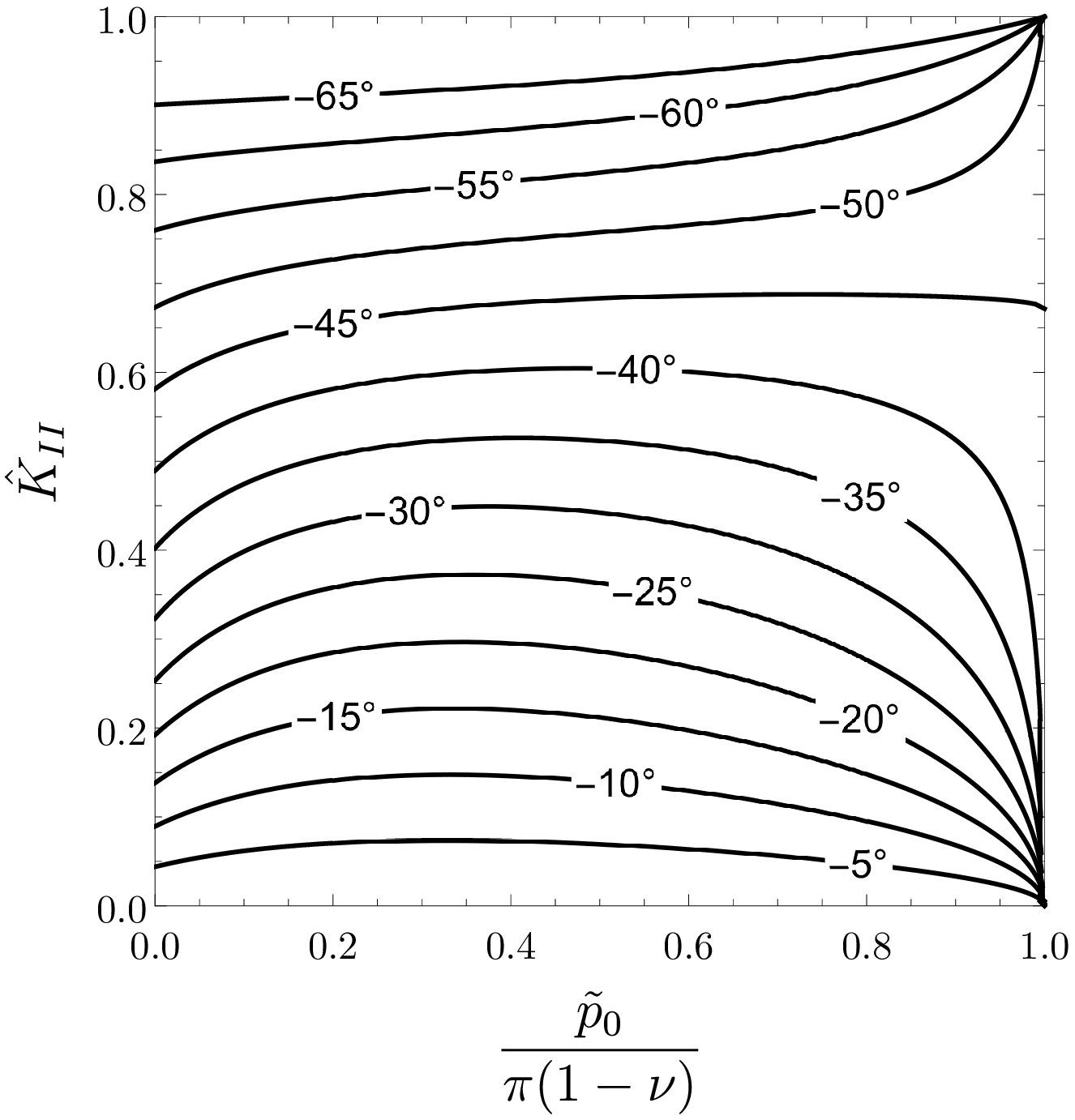}
\put(-165,20){$A$}
\put(5,20){$B$}
\put(-165,180){$D$}
\put(5,180){$C$}
\caption{MSED: Predicted propagation angle $\theta_f$ for $\hat K_{II} \in [0,1]$ and $\tilde p_0 \in [0,\pi(1-\nu)]$ for $\nu=0.3$.}
\label{MSED_n03}
\end{center}
\end{figure}

We now analyse the fracture propagation angle $\theta_f$, as computed in \eqref{tf_S}, for $\nu=0.3$ and all admissible values of $\hat K_{II} \in [0,1]$ and $\tilde p_0 \in [0,\pi(1-\nu)]$. The corresponding results are presented in Fig. \ref{MSED_n03}.

%%%%%%%%%%%%%%%%%%%%%%%%%%%%%%%%%%%%%%%%%%%%%%%%%%%%%%%%%%%%%%%%%%%%%%%%
%%%%%%%%%%%%%%%%%%%%%%%%%%%%%%%%%%%%%%%%%%%%%%%%%%%%%%%%%%%%%%%%%%%%%%%%

On the edge AB ($K_{II}=0$) the angle of crack propagation is $\theta_f=0$. Furthermore, for pure Mode II ($\hat K_{II}=1$ or edge CD) the solution can be found analytically:
\begin{equation}\label{msed_mode_II}
\theta_f=-\arctan\frac{2\sqrt{2+\nu-\nu^2}}{1-2 \nu}.
\end{equation}
We recall that for the MCS criterion, the corresponding result was $\theta_f=-2 \arctan (1/\sqrt{2})$. However, the biggest difference with MCS appears along the edge $BC$ (viscosity dominated regime, $\hat K_I=0$), where the redirection angle was previously equal to zero.

\begin{figure}[htb!]
\begin{center}
\includegraphics[scale=0.7]{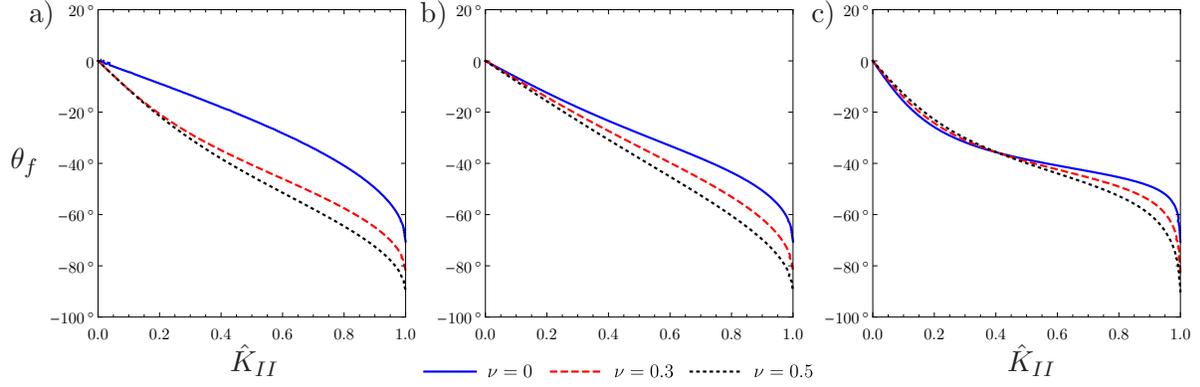}
\put(-447,80){$\theta_f$}
\put(-364,3){$\hat K_{II}$}
\put(-70,3){$\hat K_{II}$}
\put(-440,135){a)}
\put(-292,135){b)}
\put(-144,135){c)}
\caption{MSED: Redirection angle, $\theta_f$, for various values of Poisson's ratio and: a) $\frac{\tilde p_0}{\pi(1-\nu)}=0$, b) $\frac{\tilde p_0}{\pi(1-\nu)}=0.5$, c) $\frac{\tilde p_0}{\pi(1-\nu)}=0.9$.}
\label{MSED_nu}
\end{center}
\end{figure}

\begin{figure}[htb!]
\begin{center}
\includegraphics[scale=1.1]{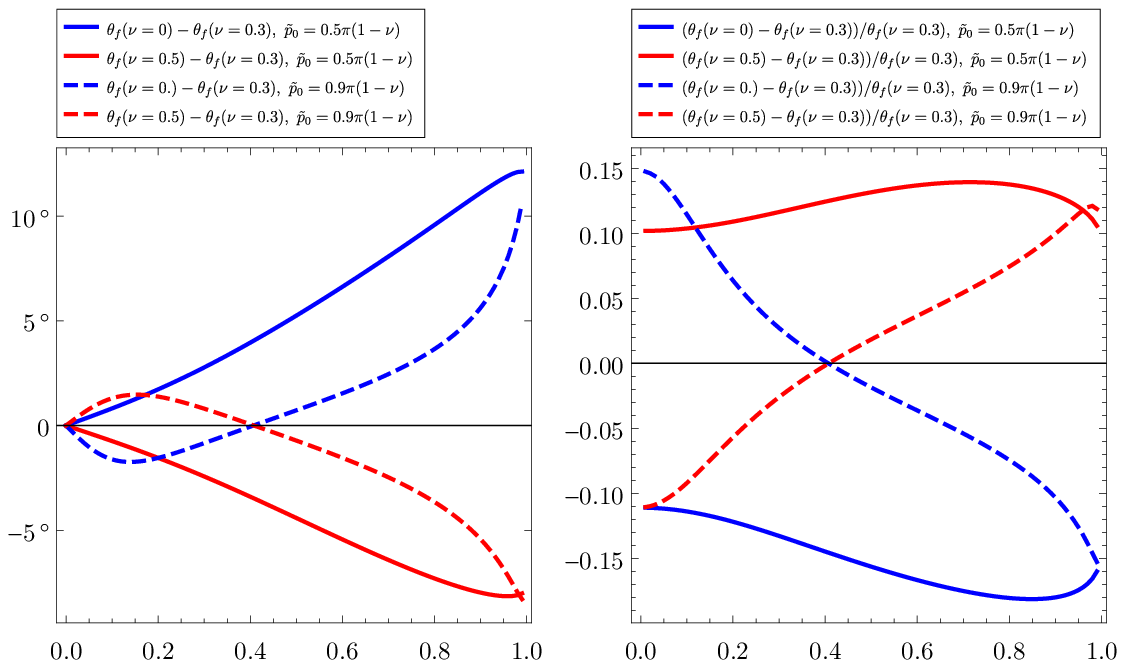}
%\put(-430,66){$\theta_f$}
\put(-268,-12){$\hat K_{II}$}
\put(-86,-12){$\hat K_{II}$}
\put(-360,195){a)}
\put(-172,195){b)}
\caption{MSED:
Absolute (a) and relative (b) deviations of the redirection angle, $\theta_f$, from a reference value obtained in the case of $\nu=0.3$ for two limiting values of the Poisson's ratio: $\nu=0$ and $\nu=0.5$. }
\label{MSED_nu_diff}
\end{center}
\end{figure}

In Fig. \ref{MSED_nu} -- Fig. \ref{MSED_nu_diff}, we show the dependence of $\theta_f$ on the Poisson's ratio. The impact of $\nu$ is much more pronounced here than in the case of the MCS criterion. The discrepancies between the respective results increase with increasing $\hat K_{II}$. Moreover, for $\tilde p_0=0$, the difference between the angles obtained for $\nu=0$ and $\nu=0.5$ is the greatest, amounting to a maximum of $12^\circ$ (see Fig. \ref{MSED_nu_diff}).

%%%%%%%%%%%%%%%%%%%%%%%%%%%%%%%%%%%%%%%%%%%%%%%%%%%%%%%%%%%%%%%%%%%%%%%%
%%%%%%%%%%%%%%%%%%%%%%%%%%%%%%%%%%%%%%%%%%%%%%%%%%%%%%%%%%%%%%%%%%%%%%%%

%%%%%%%%%%%%%%%%%%%%%%%%%%%%%%%%%%%%%%%%%%%%%%%%%%%%%%%%%%%%%%%%%%%%%%%%
%%%%%%%%%%%%%%%%%%%%%%%%%%%%%%%%%%%%%%%%%%%%%%%%%%%%%%%%%%%%%%%%%%%%%%%%

\clearpage

\section{Conclusions}
\label{sec:conclusions}

In the framework of classical Linear Elastic Fracture Mechanics, the existing criteria for determination of the deflection angle
for a small kink give similar results. All of them utilise the asymptotic analysis of the strain-stress fields in the near-tip zone, and the obtained redirection angles usually depend on the relationship between the stress intensity factors (see, for example, the here discussed MSC and MSED criteria for $\tilde p_0=0$).

%% MARKER %%

We showed that accounting for the hydraulically induced tangential traction on the fracture walls, by introducing one more component of loading, changes the corresponding results with respect to those predicted by the  classical criteria. The greatest discrepancies are obtained in the case of substantial external shear load ($\hat K_{II}\to1$), which occurs when the position of the initial crack does not match the orientation of the principal stresses for small material toughness (while approaching the so-called viscosity dominated regime $\tilde p_0\to\pi(1-\nu)$). In such a situation the crack redirection angle is extremely sensitive to the values of both, $\hat K_{II}$ and $\tilde p_0$.

The criteria analysed in this paper, MCS and MSED, exhibit different sensitivity to the value of Poisson's ratio. Clearly, the predictions made here need to be verified experimentally, which constitutes a real technical challenge.

Finally, other classical criteria for the fracture redirection should be revisited when considering the problem of a fluid driven crack.

\section*{Acknowledgments}
The authors gratefully acknowledge financial support from the European Union's Seventh Framework Programme FP7/2007-2013/ under REA Grant agreement numbers: PCIG13-GA-2013-618375-MeMic (A.P.), PITN-GA-2013-606878-CERMAT2 (M.P.) and IRSES-GA-2013-610547-TAMER (M.W.). One of the authors, G. Mishuris, thanks for a support during his visit to Russia by grant 14.Z50.31.0036 awarded to R. E. Alexeev Nizhny Novgorod Technical University by Department of Education and Science of the Russian Federation. Finally, the authors are thankful to Prof. M. Kachanov for fruitful discussions and useful comments.

\appendix
\renewcommand{\theequation}{\thesection.\arabic{equation}}

\section{Functions ${\bf \Phi}_j(\theta)$ and ${\bf \Psi}_j(\theta)$
equations \eqref{eq:asymu} and \eqref{eq:asyms}}
\setcounter{equation}{0}
\label{app1}
\[
\Phi_{I}^r(\theta) =
\frac{1+\nu}{E} \cos \frac{\theta }{2} \left[3-4\nu- \cos \theta \right],
\]
\[
\Phi_{I}^\theta(\theta) =
-\frac{1+\nu}{E}\sin \frac{\theta }{2} \left[3-4\nu- \cos \theta \right],
\]
\[
\Phi_{II}^r(\theta) =
\frac{1-\nu^2}{E} \sin \frac{\theta }{2} \left[\frac{3\nu}{1-\nu} -1+\frac{3\cos \theta}{1-\nu}\right],
\]
\[
\Phi_{II}^\theta(\theta) =
-\frac{1-\nu^2}{E} \cos \frac{\theta }{2} \left[5+\frac{\nu}{1-\nu}-\frac{3\cos \theta}{1-\nu} \right],
\]
\[
\Phi_{III}^z(\theta) = \frac{4(1+\nu)}{E}\sin \frac{\theta }{2},
\]
\[
\Phi_{\tau}^r(\theta) = -\frac{4\left(1-\nu^2 \right)}{E}  \cos \frac{3 \theta}{2},\quad
\Phi_{\tau}^\theta(\theta) = \frac{4 \left(1-\nu^2 \right) }{E}\sin \frac{3 \theta}{2}.
\]
For the plane strain $\Psi^{zz} =\nu \left(\Psi^{rr}+\Psi^{\theta\theta} \right)$.
\[
\Psi_{I}^{rr}(\theta) = \frac{1}{4}\left[5\cos \frac{\theta }{2} - \cos \frac{3 \theta}{2}\right],\quad
\Psi_{I}^{\theta\theta}(\theta) = \cos^3\frac{\theta}{2},
\]
\[
\Psi_{I}^{r\theta}(\theta) =
\frac{1}{2}\cos\frac{\theta}{2}\sin\theta,\quad
\Psi_{II}^{rr}(\theta) =
-\frac{1}{4}\left[5\sin \frac{\theta }{2} -3 \sin \frac{3 \theta}{2}\right],
\]
\[
\Psi_{II}^{\theta\theta}(\theta) =-3\sin\frac{\theta}{2}\cos^2\frac{\theta}{2},\quad
\Psi_{II}^{r\theta }(\theta) =\frac{1}{4}\left[\cos\frac{\theta}{2}+3\cos\frac{3\theta}{2}\right],
\]
\[
\Psi_{III}^{r z }(\theta) =\sin\frac{\theta}{2},\quad
\Psi_{III}^{\theta z}(\theta) =\cos\frac{\theta}{2},
\]
\[
\Psi_{\tau}^{rr}(\theta) =-\Psi_{\tau}^{\theta\theta}(\theta)=-2(1-\nu)\cos \frac{3 \theta}{2},\quad
\Psi_{\tau}^{r\theta}(\theta) =
2(1-\nu)\sin\frac{3\theta}{2}.
\]

\end{document}